\documentstyle[preprint,aps]{revtex}
\begin{document}
\setcounter{equation}{0}
\setcounter{section}{0}
\renewcommand{\thesection}{\arabic{section}}
\renewcommand{\theequation}{\thesection.\arabic{equation}}
\preprint{Preprint No. IP--BBSR--95/76}
\title{An Auxiliary 'Differential Measure' for $SU(3)$}
\author{J. S. Prakash{\thanks {jsp@iopb.ernet.in}}} 
\address{Institute Of Physics\\ Sachivalaya Marg, Bhubaneswar 751 005, India}
\date{April 1996}
\maketitle
\begin{abstract} 
A 'differential measure' is used to cast our calculus for the
group $SU(3)$ into a form similar to Schwinger's boson operator
calculus for the group $SU(2)$.  It is then applied to compute
(i) the inner product between the basis states and (ii) an
algebraic formula for the Clebsch-Gordan coefficients.  These
were obtained earlier by us using Gaussian integration
techniques.
\end{abstract}
PACS number(s): 
\pacs{}

\newpage

\section{\bf Introduction}

In a previous paper\cite{JSPHSS} we set up a calculus for
dealing with computations in $SU(3)$.  One of the main
ingredients of this calculus is an 'auxiliary' Gaussian measure
using which all computations on $SU(3)$ are reduced to Gaussian
integrations.  As an application of this calculus we also
derived a closed form algebraic expression for the
Clebsch-Gordan coefficients involving direct products of
arbitrary irreducible reprsentations of $SU(3)$.  Apart from
solving this important and long standing problem completely this
calculus can also be used advantageously in any situation
involving computations over irreducible representations of
$SU(3)$.  Therefore it wolud be interesting to develop
alternative ways of implementing the tools of this calculus. 

In this paper we use an 'auxiliary differential measure' in
place of the above mentioned Gaussian measure and obtain all our
previous results for $SU(3)$.  In contrast to our erlier scalar
product which involved integrations using a Gaussian measure
here the 'differential measure' involves a differential operator
which operates on functions of basis states and gives the scalar
product as the end product.  This present method corresponds to
the operator techniques in field theories as opposed to the path
integral techniques of our erlier method.  Therefore this is not
a different inner product but is rather a different technique of
evaluating the same inner product which we used in our erlier
paper.  A similar measure was used by Schwinger\cite{SJ} in his
systematic derivation of the results of the angular momentum
algebra.  Later Ruegg\cite{RH} used a simplified version of it
to obtain the basis states and Clebsch-Gordan coefficients of
$SU_q(2)$.  We have retained the name 'differential measure',
first used by Ruegg in the reference cited, to denote the
differntial operator that we mentioned in the above.  We make
use of Schwinger's evaluation of the scalar products fully and,
for this purpose, reproduce his derivation in the appendix to
this paper.

The plan of the paper is as follows.  In section 2 we review our
previous results which we obtained using our calculus for
$SU(3)$.  Then in section 3 we derive the auxiliary
normalizations of our basis states for $SU(3)$ using an
'auxiliary differential measure' and show that it agrees with
our previous results using an 'auxiliary' Gaussian measure.
Next in section 4 we apply the 'differential measure' to obtain
an algebraic formula for the Clebsch-Gordan coefficients of
$SU(3)$ and show that our results in this case also agree with
our previous ones.  In section 5 we derive the overall
normalizations of the Clebsch-Gordan coefficients.  This
corresponds to computing the norms of the invarints.  In our
previous work we have left this to be computed by hand.  The
last section is devoted to a discussion of the results of this
paper.

\setcounter{equation}{0}
\section{Overview of our previous results}

$SU(3)$ is the group of $3\times 3$ unitary unimodular matrices
$U$ with complex coefficients. It is a group of $8$ real
parameters.  The matrix elements satisfy the following
conditions

\begin{eqnarray}
U&=&(u_{ij}), \nonumber \\ 
U^\dagger U &=& I, \qquad UU^\dagger = I\, ,\,\,\,
\mbox{where}\,\, I \,\,  \mbox{is the identity
matrix and\, ,}\nonumber \\ 
\mbox{det}(U)&=&1\, .
\end{eqnarray}

\subsection{ Parametrization.}

One well known parametrization of $SU(3)$ is due to Murnaghan\cite{MFD}.
In this we write a typical element of $SU(3)$ as :

\begin{equation}
D(\delta_1,\delta_2,\phi_3) U_{23}(\phi_2,\sigma_3) U_{12}(\theta_1,\sigma_2) 
U_{13}(\phi_1,\sigma_1)\, ,
\end{equation}

with the condition $\phi_3 = -(\delta_1 + \delta_2)$.  Here $D$
is a diagonal matrix whose elements are $\exp (i\delta_1)$,
${\exp}(i\delta_2)$ ,${\exp}(i\phi_3)$ and $U_{pq}(\phi,\sigma)$ is a $3
\times 3$ unitary unimodular matrix which for instance in the
case $p=1$, $q=2$ has the form

\begin{equation}
\pmatrix{\cos\phi               & -\sin\phi {\exp}(-i\sigma) & 0 \cr
         \sin \phi {\exp}(i\sigma) & \cos \phi               & 0 \cr
         0                     & 0                      & 1}\, .
\end{equation}
The $3$ parameters $\phi_1$, $\phi_2$, $\phi_3$ are longitudinal
angles whose range is $-\pi \leq \phi_i \leq \pi$; and the
remaining $6$ parametrs are latitude angles whose range is
$\-\frac{1}{2} \pi \leq \sigma_i \leq \frac{1}{2}\pi$.

Now the transformations $U_{23}$ and $U_{13}$ can be changed into
transformations of the type $U_{12}$ whose matrix elements are
known, by the following device

\begin{eqnarray}
U_{13}(\phi_1, \sigma_1) &=& (2,3) U_{12}(\phi_1,\sigma_1) (2,3), \nonumber \\
U_{23}(\phi_2, \sigma_3) &=& (1,2) (2,3) U_{12}(\phi_2,\sigma_3)
(2,3) (1,2)\, ,
\end{eqnarray}

where $(1,2)$ and $(2,3)$ are the transposition matrices

\begin{equation}
(1,2) = \pmatrix{0 & 1 & 0 \cr
                 1 & 0 & 0 \cr
                 0 & 0 & 1 }, \hspace{.1in} (2,3) = \pmatrix{0 & 1 & 0 \cr
                                                      1 & 0 & 0 \cr
                                                      0 & 0 & 1 }\, .
\end{equation}

In this way the expression for an element of the $SU(3)$ group becomes

\begin{equation}
D(\delta_1,\delta_2,\phi_3) (1,2) (2,3) U_{12}(\phi_2,\sigma_3)
(2,3) (1,2) U_{12}(\theta_1,\sigma_2) (2,3)
U_{12}(\phi_1,\sigma_1) (2,3)\, .
\end{equation}

\subsection{Irreducible representations.}

The parametrization, described in the previous section, provides
us with a defining irreducible representation $\underline{3}$ of
$SU(3)$ acting on a $3$ dimensional complex vector space spanned
by the triplet $z_1, z_2, z_3$ of complex variables.  The
hermitian adjoint of the above matrix gives us another defining
but inequivalent irreducible representaion $\underline{3^*}$ of
$SU(3)$ acting on the triplet ${w_1,w_2,w_3}$ of complex
variables spanning another $3$ dimensional complex vector space.
Tensors constructed out of these two $3$ dimensional
representations span an infinite dimensional complex vector
space\cite{HB}.  

\subsection{{{The Constraint}}}

If we impose the constraint

\begin{eqnarray}
z_1w_1+z_2w_2+z_3w_3=0\, ,
\label{z.w}
\end{eqnarray}
on this space we obtain an infinite dimensional complex vector
space in which each irreducible representation of $SU(3)$ occurs
and once and only once.  Such a space is called a model space
for $SU(3)$.  Further if we solve the constraint
$z_1w_1+z_2w_2+z_3w_3=0$ and eliminate one of the variables, say
$w_3$, in terms of the other five variables $z_1, z_2, z_3, w_1,
w_2$ we can write a genarating function to generate all the
basis states of all the IRs of $SU(3)$ each exactly only once.
This generating function is computationally a very convenient
realization of the basis of the model space of $SU(3)$.
Moreover we can define an auxiliary scalar product on this space
by choosing one of the variables, say $z_3$, to be a planar
rotor ${\exp}(i\theta)$.  Thus the model space for $SU(3)$ is
now a Hilbert space with this scalar product between the basis
states.  But it must be admitted that this scalar product is not
$SU(3)$ invariant though the normalizations obtained from it can
be easily related to the ones got from a group invariant scalar
product.

The above construction was carried out in detail in a previous
paper by us\cite{JSPHSS}.  For easy accessability we give a
self-contained summary of those results here.

\subsection{{{Labels for the basis vectors.}}}

\noindent {\bf{(i). Gelfand-Zetlein labels}}\\

Normalized basis vectors are denoted by,
$\vert{M,N;P,Q,R,S,U,V}>$.  All labels are non-negative
integers.  All Irreducible Represenatations(IRs) are uniquely
labeled by $(M, N)$.  For a given IR $(M, N)$, labels
$(P,Q,R,S,U,V)$ take all non-negative integral values subject to
the constraints:

\begin{equation}
R+U=M\hspace{.1in},\hspace{.1in} S+V=N\hspace{.1in},\hspace{.1in} P+Q=R+S\, .
\end{equation}

The allowed values can be prescribed easily: $R$ takes all values
from $0$ to $M$, and $S$ from $0$ to $N$.  For a given $R$ and
$S$, $Q$ takes all values from $0$ to $R+S$.\\

\noindent {{\bf{(ii). Quark model labels}}}.\\

The relation between the Gelfand-Zetlein labels and the
Quark Model labels is as given below.

\begin{eqnarray}
2I=P+Q=R+S, \,\, 2I_3=P-Q, \,\, Y = \frac{1}{3} (M-N) + V-U 
= \frac{2}{3} (N-M)-(S-R)\, .\nonumber\\
\end{eqnarray}

\subsection{\bf{{Explicit realization of the basis states.}}}

\noindent {\bf{(i). {Generating function for the basis states of $SU(3)$}}}

The generating function for the basis states of the IR's of
$SU(3)$ can be written as

\begin{equation}
g(p,q,r,s,u,v)={\exp}(r(pz_1+qz_2)+s(pw_2-qw_1)+uz_3+vw_3)\, .
\label{g()}
\end{equation}

The coefficient of the monomial $p^Pq^Qr^Rs^Su^Uv^V$ in the
Taylor expansion of Eq.(\ref{g()}) in terms of these monomials
gives the basis state of $SU(3)$ labelled by the quantum numbers
$P, Q, R, S, U, V$. 

\noindent {\bf{{ (ii). Formal generating function for the basis
states of $SU(3)$ }}}

The generating function Eq.(\ref{g()}) can be written formally
as 

\begin{equation}
g=\sum_{P,Q,R,S,U,V} p^Pq^Qr^Rs^Su^Uv^V \vert PQRSUV)\, ,
\end{equation}
where $\vert PQRSTUV)$ is an unnormalized basis state of $SU(3)$
labelled by the quantum numbers $P,Q,R,S,U,V$.

Note that the constraint $P+Q=R+S$ is automatically satisfied in
the formal as well as explicit Taylor expansion of the generating
function.

\noindent {\bf{{ (iii). Generalized generating function for the basis
states of $SU(3)$}}}

It is useful, while computing the normalizations of the basis
states, to write the above generating function in the following
form
\begin{equation}
{\cal G}(p,q,r,s,u,v)={\exp}(r_pz_1+r_qz_2+s_pw_2+s_qw_1+uz_3+vw_3)\, .
\label{GGF}
\end{equation}

In the above generalized generating function (\ref{GGF}) the
following notation holds.

\begin{equation}
r_p=rp, \qquad r_q=rq, \qquad s_p=sp, \qquad s_q=-sq\, .
\label{s_q=-sq}
\end{equation}

\subsection{Notation}
In the recapitulation below and in the derivations later we will
assume that all our variables (except the $z_i$ and the $w_i$ )
are real (eventhough we have treated them as complex variables
at some places) since we are interested only in extracting the
various objects, auxiliary normalization constants of basis
states, Clebsch-Gordan coefficients etc., occuring as
coefficients in the expansion of the scalar products as power
series in these variables.  

\subsection{'Auxiliary' scalar product for the basis functions.}

The scalar product to be defined in this section is 'auxiliary'
in the sense that it does not give us the 'true' normalizations
of the basis astes of $SU(3)$.  However it is computationally
very convenient for us as all computations with this scalar
product get reduced to simple Gaussian integrations and the
'true' normalizations themselves can then be got quite easily.  

{\bf{(i). Scalar product between generating functions of basis
states of $SU(3)$}}

We define the scalar product between any two basis states in
terms of the scalar product between the corresponding generating
functions as follows :

\begin{eqnarray}
(g', g)&=& {\int_{-\pi}^{+\pi}}{\frac{d\theta}{2\pi}} \int
\frac{d^{2}z_1}{\pi^2} \frac{d^{2}z_2}{\pi^2}
\frac{d^{2}w_1}{\pi^2} \frac{d^{2}w_2}{\pi^2}
{\exp}(-\bar{z_1}z_1 - \bar{z_2}z_2 - \bar{w_1}w_1
-\bar{w_2}w_2)\nonumber\\
&&\nonumber \\
&&\times {\exp}((r'(p'z_1+q'z_2) + s'(p'w_2-q'w_1) - \frac{-v'}{z_3}
(z_1w_1 + z_2w_2) + u'\bar{z}_3) \nonumber \\
&&\nonumber\\
&&\times {\exp}((r(pz_1 + qz_2) + s(pw_2-qw_1) - \frac{-v}{z_3}
(z_1w_1 + z_2w_2) + uz_3)\, , \nonumber \\
&&\nonumber \\
&=& (1-v'v)^{-2} \left (\sum_{n=0}^{\infty}
\frac{(u'u)^n}{(n!)^2}\right )
{\exp}\left [(1-v'v)^{-1}(p'p + q'q)(r'r + s's)\right ]\, .
\label{gg'}
\end{eqnarray}

{\bf{(ii). Choice of the variable $z_3$}}

To obtain the Eq.(\ref{gg'} we have made the choice
\begin{eqnarray}
z_3=\exp(i\theta )\, .
\label{z3}
\end{eqnarray}

The choice, Eq.(\ref{z3}), makes our basis states for $SU(3)$
depened on the variables $z_1,z_2,w_1,w_2$ and $\theta $.

{\bf{(iii). Scalar product between the gneralized generating
functions of the basis states of $SU(3)$}} 

For the generalized generating function the scalar product
becomes 

\begin{eqnarray}
{(\cal{G'}, \cal{G})} &=& (1 - v'v)^{-2} 
{\exp}\left [(1 - v'v)^{-1}({r_p}'r_p +
{r_q}'r_q + {s_p}'s_p + {s_q}'s_q) \right ]\nonumber \\
&&\nonumber \\
&&\times \left [{\sum_{n=0}}^{\infty} \frac{1}{(n!)^2}\left (u'- v
\frac{({r_p}'{s_q}'+ {r_q}'{s_p}' )}{(1 - v'v)}\right )^n\, {\bf
\cdot}\, \left (u - v'\frac{({r_p}{s_q} + {r_q}{s_p})}{(1 -
v'v)}\right )^n \right ]\, .
\label{GGFSP}
\end{eqnarray}

If in this we substitute from Eq.(\ref{s_q=-sq}) for $r_p,
\cdots , s_q$  etc we get back the scalar product Eq.(\ref{gg'}) 
between the ordinary generating functions Eq.(\ref{g()}).

\subsection{Normalizations}
Based on the scalar products defined in the previous subsection
we arrive at the normalizations described below.

\noindent {\bf{{(a). 'Auxiliary' normalizations of unnormalized
basis states}}} 

The scalar product between two unnormalized basis states,
computed using our 'auxiliary scalar product, is given by

\begin{eqnarray}
M(PQRSUV)&\equiv &(PQRSUV\vert PQRSUV)\, ,\nonumber \\
&&=\frac{(V+P+Q+1)! }{P! Q! R!S! U! V! (P+Q+1)}\, .
\label{M}
\end{eqnarray}

\noindent {\bf{(b). Scalar product between the unnormalized and normalized
basis states}}

The scalar product, computed using our 'auxiliary' scalar
product, between an unnormalized basis state and a normalized
one is denoted by $(PQRSUV\Vert PQRSUV>$ and is given below

\begin{equation}
(PQRSUV\Vert PQRSUV>=N^{-1/2}(PQRSUV)\times M(PQRSUV)\, .
\label{(||)}
\end{equation}

\noindent {\bf{{(c). 'True' normalizations of the basis
states}}} 

We call the ratio of the 'auxiliary' norm of the unnormalized
basis sate represented by $\vert PQRSUV)$ and the scalar product
of the normalized Gelfand-Zeitlin state, represented by $\vert
PQRSUV > $ with an unnormalized basis state, as 'true'
normalization.  It is given by

\begin{eqnarray}
N^{1/2}(PQRSUV)&\equiv & \frac{(PQRSUV\Vert PQRSUV)} 
{(PQRSUV\Vert PQRSUV>}\, ,\nonumber \\
&&=(\frac{(U+P+Q+1)! (V+P+Q+1)! }{P! Q! R!S! U! V! (P+Q+1)})^{1/2}\, .
\end{eqnarray}

\subsection{Generating function for the invariants.\label{GFI}}

All Clebsch-Gordan coefficients can be extracted from the
following generating function of the invariants :

\begin{eqnarray}
I_{\pm}(j_{12}, j_{23}, j_{31}, j_{21}, j_{32}, j_{13},
j_{\pm}) &=& {\exp}(j_{12}\vec{z^1}\cdot\vec{w^2} +
j_{23}\vec{z^2}\cdot\vec{w^3} + j_{31}\vec{z^3}\cdot\vec{w^1}
+j_{21}\vec{z^2}\cdot\vec{w^1}\nonumber\\
&& + j_{32}\vec{z^3}\cdot\vec{w^2}
+ j_{13}\vec{z^1}\cdot\vec{w^3}+j_+{\vec z}^1\cdot{\vec
z}^2\times {\vec z}^3\,\, \nonumber \\
&&~~~~~~~~~~~~~~~~~~~~~~~~~~~~
 {\mbox {\bf or}}\,\, j_-{\vec w}^1\cdot{\vec
w}^2\times {\vec w}^3)\, .
\end{eqnarray}
where the $j_{mn}, m, n=1,2,3, m\neq n$ and $j_+, j_-$ are any
eight complex variables.

\subsection{Multiplicity labels for the Clebsch-Gordan series.}

For given three IRs, $(M^{1},N^{1}), (M^{2},N^{2}),
(M^{3},N^{3})$, construct all the solutions of

\begin{eqnarray}
N(1,2) + N(1,3) + L\epsilon(L) &=& M^{1} {\, ,\nonumber}\\ 
N(2,3) + N(2,1) + L\epsilon(L) &=& M^{2} {\, ,\nonumber} \\
N(3,1) + N(3,2) + L\epsilon(L) &=& M^{3}{\, ,\nonumber} \\
N(2,1) + N(3,1) + \vert{L}\vert\epsilon(-L) &=& N^{1}{\, ,\nonumber} \\
N(3,2) + N(1,2) + \vert{L}\vert\epsilon(-L) &=& N^{2}{\, ,\nonumber} \\
N(1,3) + N(2,3) + \vert{L}\vert\epsilon(-L) &=& N^{3}{\, ,\nonumber} \\
3L &=& \sum^{3}_{n=1}(M^a-N^a)\, ,
\end{eqnarray}
where $N(a,b), a = b$ are non-negative integers.  They provide
unambiguous labels for the Clebsch-Gordan series as follows.
For given two IRs $(M^{a},N^{a})$ and $(M^{a'},N^{a'})$,
construct all $(M^{a"},N^{a"})$ for which $N(a,b), a\neq{b}$ have
non-negative integer solutions.  Then the reversed pair
$(N^{a"},M^{a"})$ gives all IRs in the Clebsch-Gordan series.
Multiplicity of solutions for one $(M^{a"},N^{a"})$ provides the
multiplicity of repeating IRs.  Therefore $N(a,b)$ unambiguously
provide the multiplicity labels.

\subsection{Clebsch-Gordan coefficients.}

\noindent (i). {\bf{{$3-SU(3)$ Symbol.}}}\\

$3-G$ symbols are related\cite{JSPHSS} to the Clebsch-Gordan
coefficients and have more explicit symmetry than the latter.
The $3-SU(3)$ symbol is represented by,

\begin{equation}
{\tiny{
\left [
\matrix{& N(1,2) && N(1,3) && N(2,3) && N(2,1) && N(3,1) && N(3,2) \cr
N^2 && M^1 && N^3 && M^2 && N^1 && M^3 && N^2\cr
&& V^1, R^1 && V^3, S^3 && V^2, R^2 && V^1, S^1 && V^3, R^3 && V^2, S^2 \cr
&& P^1 && P^3 && P^2 && Q^1 && Q^3 && Q^2}
\right ]_L\, .
}}
\end{equation}

Here the top row specifies the multiplicity labels.  The second
and third rows specify the usual complete set of labels for the
basis states of the three IRs.\\

\noindent (ii). {{\bf{Generating function for the $3-SU(3)$ 
symbol for $L > 0$.}}}\\

Extract the coefficient of the monomial

\begin{equation}
j_{12}^{N(1,2)}j_{13}^{N(1,3)}j_{31}^{N(3,1)}j_{23}^{N(2,3)}j_{32}^{N(3,2)}
j_+^L\prod_{\alpha=1}^3 \bar p^{P_\alpha}\bar q^{Q_\alpha}\bar
r^{R_\alpha}\bar u^{U_\alpha}\bar v^{V_\alpha}\, ,\nonumber
\end{equation}

in the expansion of 

\begin{eqnarray}
&&
\Vert B \Vert^2
{\exp}  \left [ \Vert B \Vert ((j_+ r^1  r^2 u^3 
- r^1j_{12} s^2 +  r^2j_{21} s^1 +  r^1j_{12} s^2( u^2j_{23} v^3
+ u^1j_{13} v^3)\right. \nonumber\\
&&{{- r^2j_{21} s^1( u^2j_{23} v^3+ u^2j_{23} v^3)
+ r^1j_{13} v^3 u^3j_{32} s^2 - r^2j_{23} v^3 u^3j_{31} s^1)}}
  \times (p^1 q^2 -  p^2 q^1) \nonumber \\
&&\left. + (cyclic) \right ]\, ,
\end{eqnarray}
where
\begin{eqnarray}
\Vert B^{-1}\Vert &=& det(1-{\bar J}V)\, ,\\
&&\nonumber\\
{\bar{V}}&=&{\pmatrix{{\bar {v}}^1 & 0             & 0\cr 
                                 0 & {\bar {v}}^2  & 0\cr
0 & 0 & {\bar {v}}^3}} \, ,\nonumber \\
&& \nonumber \\
&& \nonumber \\
J&=&{\pmatrix{ j_{31}+j_{21} & -j_{12}        & -j_{13} \cr 
              -j_{21}        &  j_{12}+j_{32} & -j_{23} \cr 
              -j_{31}        & -j_{32}        &  j_{23}+j_{13} }}\, .
\label{V&J}
\end{eqnarray}
and multiply it by the factor 

\begin{equation}
\prod_{\alpha=1}^3\left [ {\frac{P^\alpha !Q^\alpha !R^\alpha !S^\alpha
!U^\alpha !V^\alpha !(U^\alpha !+2I^\alpha !1)(2I^\alpha !+1)}
{(V^\alpha !+2I^\alpha !+1)}}\right ]^{1/2}\, .
\end{equation}

This gives the $3-SU(3)$ symbol up to an overall normalization
depending only on IRs involved.\\

\noindent (iii). {{\bf{Formula for $3-SU(3)$ symbol for $L > 0$.}}}\\

We have obtained an explicit analogue of the Bargamann's formula
for the $3-j$ symbol of $SU(2)$.  This formula for $L > 0$ is given below
\begin{eqnarray}
\left [
\matrix{& N(1,2) && N(2,3) && N(3,1) && L && N(1,3) && N(3,2) & N(2,1) \cr
&&& M^1 N^1 &&&& M^2 N^2 &&&& M^3 N^3 \cr
&&& I^1 I^1_3 Y^1 &&&& I^2 I^2_3 Y^2 &&&& I^3 I^2_3 Y^3}
\right ]\nonumber\\
\end{eqnarray}

\begin{eqnarray}
&=&n(N(1,2), N(2,3), N(3,1), N(2,1), N(3,2), N(1,3), L)\nonumber \\
&&\nonumber \\
&&\times  \prod_{\alpha=1}^3 \left [ {
\frac{P^\alpha !Q^\alpha !R^\alpha !S^\alpha
!U^\alpha !V^\alpha !(U^\alpha !+2I^\alpha !1)(2I^\alpha !+1)}
{(V^\alpha !+2I^\alpha !+1)}} \right ]^{1/2}\nonumber \\
&&\nonumber \\
&&\times  \sum_{e,f,g,k,l,m}{\frac{(1+\sum e()+f()+g()+k()+l()+m()+n()!)}
{(1+\sum (k()+l()+m()+n()! \prod l()!...n()!)}}\nonumber\\
&&\nonumber \\
&&\times (-1)^{\sum_s (k()+n())+\sum_Am()+\sum f()+g())}\, ,
\label{AECG}
\end{eqnarray}
where $e(), f(), g(), l(), k(), m(), n()$ are the powers of the
various monomials occuring in the above expansion.  The notation
is explained in the tables given below.  The symbol 
\begin{eqnarray}
n(N(1,2), N(2,3), N(3,1), N(2,1), N(3,2), N(1,3), L)\, ,
\label{overall}
\end{eqnarray}
stands for the overall normalization of the Clebsch-Gordan
coefficients and therefore it does not depend on any particular
IR.  It is the norm of the invariants\cite{BV}.

\begin{eqnarray}
P^\alpha &=& \sum (l(\alpha --)+k(\alpha ---)+m(\alpha
-----)+n(\alpha -----));\nonumber \\ Q^\alpha &=& \sum
(l(-\alpha -)+k(-\alpha --)+m(-\alpha ----)+n(-\alpha
----));\nonumber \\ R^\alpha &=& \sum (l(\alpha --)+l(-\alpha
-)+k(--\alpha -)+m(--\alpha ---)+n(--\alpha ---));\nonumber \\
S^\alpha &=& \sum (k(---\alpha )+m(---\alpha --)+n(---\alpha
--));\nonumber \\ U^\alpha &=& \sum (l(--\alpha )+m(----\alpha
-)+\!n\!(----\alpha -)+\!e\!(-\alpha
-)+\!f\!(\alpha---)+\!f\!(--\alpha -) \nonumber \\
&&+2g(\alpha\beta\gamma));\nonumber \\ V^\alpha &=& \sum
(m(-----\alpha )+n(-----\alpha )+e(-\alpha )+f(-\alpha
--)+f(---\alpha)+g(-\alpha-)\nonumber \\
&&+g(--\alpha));\nonumber \\ L^\alpha
&=& \sum (---);\nonumber \\ N(\alpha, \beta ) &=& \sum
(k(--\alpha\beta )+m(--\alpha\beta --)+m(----\alpha\beta
)+n(--\alpha\beta --)+n(---\alpha\beta )\nonumber \\
&&+e(\alpha\beta
)+f(\alpha\beta --)+f(--\alpha\beta )+g(\alpha\beta -)+g(\alpha
- \beta ));
\end{eqnarray}
Here $\sum $ stands for summation over all allowed arguments in
the blank spaces.

\vspace{0.5in}

\begin{center}
\begin{tabular}{|c|c|}\hline
$\alpha ,\beta ,\gamma \cdots $=1,2 \mbox{or} 3 & \\ \hline 
l($\alpha\beta\gamma$ ) & :($\alpha\beta\gamma$) {\mbox{is a
permutation of }} (123) \\
k($\alpha\beta\gamma\delta$) & :($\alpha\beta$)  {\mbox{is same or 
transpose of}} ($\gamma\delta$)\\
m($\alpha\beta\gamma\delta\epsilon\phi$) & :({\mbox{is a 
permutation of }} (123);\\
& :($\alpha\beta$ is same or transpose of $\gamma\delta$ );\cr
& :$\phi$ is either $\gamma$ or $\delta$ .\\
n($\alpha\beta\gamma\delta\epsilon$ ) & :($\gamma\delta\epsilon$ ) 
{\mbox{is a permutation of}} (123);\\
e($\alpha\beta$ ) :$\alpha \neq \beta$ & \\
f($\alpha\beta\gamma\delta$ ) & :$\beta = \gamma$ {\mbox{and}} 
($\alpha\beta\delta$){\mbox{is a permutation of}} (123);\\
g($\alpha\beta\gamma$) & :($\alpha\beta\gamma$ ) 
{\mbox{only even permutation of}} (123)\\ \hline
\end{tabular}
\end{center}

\vspace{0.5in}

\begin{center}
\begin{tabular}{|c|c|c|c|c|}\hline
{\mbox Monomial} & $j_+\bar p^1 \bar q^2 \bar r^1 \bar r^2 \bar u^3$ &
$\bar p^2 \bar q^1 \bar r^1 \bar r^2 \bar u^3$ & $\bar p^1 \bar
q^2 \bar r^1 j_{12} \bar s^2$ & $\bar p^2 \bar q^1 \bar r^1
j_{12} \bar s^2$ \\
\hline
{\mbox Order used in label  } & $\bar p^1 \bar q^2 \bar u^3$ & $\bar p^2
\bar q^1 \bar u^3$ & $\bar p^1 \bar q^2 \bar r^1 \bar s^2$ &
$\bar p^2 \bar q^1 \bar r^1
\bar s^2$ \\
\hline
{\mbox Power} & l(123) & l(213) & k(1212) & k(2112) \\ \hline
\end{tabular}
\end{center}

\vspace{0.5in}

\begin{center}
\begin{tabular}{|c|c|c|c|}\hline
$\bar p^1 \bar q^2 \bar r^2 j_{12} \bar s^1 $ & $ \bar p^1 \bar
q^2 \bar r^1 j_{12} \bar s^2 \bar u^2 j_{23} \bar v^3 $ & $ \bar
p^1\bar q^2\bar r^1 j_{12} \bar s^2 \bar u^1 j_{13} \bar v^3 $ &
$
\bar p^1 \bar q^2 \bar r^1 j_{13} \bar v^3 \bar u^3 j_{32} 
\bar s^2 $ \\ \hline
$ \bar p^1 \bar q^2 \bar r^2 j_{21} \bar s^1 $ & $ \bar p^1 \bar
q^2 \bar r^1 j_{12} \bar s^2\bar u^2 j_{23} \bar v^3 $ & $ \bar
p^1 \bar q^2 \bar r^1 \bar s^2 \bar u^1 \bar v^3 $ & $ \bar p^1
\bar q^2 \bar r^1 \bar s^2 \bar
v^3 \bar u^3 $ \\ \hline
\end{tabular}
\end{center}

\vspace{0.5in}

\begin{center}
\begin{tabular}{|c|c|c|c|c|}\hline
$ \bar u^3 j_{31} \bar v^1 $ & $ \bar u^2 j_{21} \bar v^1 $ & $
\bar u^3 j_{31} \bar v^1 \bar u^1 j_{12} \bar v^2 $ & $ \bar u^2 j_{21}
\bar v^1 \bar u^1 j_{32} \bar v^2 $ & $ \bar u^3 j_{32} \bar v^1
\bar u^3 j_{32} \bar v^2 $ \\ \hline
$\bar u^3 \bar v^1 $ & $ \bar u^2 \bar v^1 $
& $ \bar u^3 \bar v^1 \bar u^1 \bar v^2 $ & $ \bar u^3 
\bar v^2 \bar u^2 \bar v^1 $ & $ \bar u^3 \bar u^3 \bar u^3 \bar
v^2 \bar v^1 $ \\ \hline
$ e(31) $ & $ e(21) $ & $ f(3112) $ & $ f(3221) $ & $ g(321) $ \\ \hline
\end{tabular}
\end{center}

\setcounter{equation}{0}
\section{An auxiliary 'differential measure'}

For computing the inner product between two arbitrary basis
states of $SU(3)$ we have to find out the inner product between
two generating functions, one in primed and another in unprimed
variables, of these generating functions.  This we propose to do
in the following.  

Let us first note that 

\begin{eqnarray}
g(p,&q&,r,s,u,v; z_1,z_2,z_3,w_1,w_2) \nonumber \\ 
&=& {\exp} \left \{ r(pz_1+qz_2)+s(pw_2+qw_1)+uz_3-\frac{v}{z_3}
(z_1w_1+z_2w_2)
\right \}\nonumber \\
&=& {\exp} \left \{ p{\pmatrix{r & s}}{\pmatrix{z_1\cr w_2}}+q{\pmatrix{r &
s}}{\pmatrix{z_2 \cr -w_1}}+{\frac{v}{z_3}}\left [ {\pmatrix{z_1 &
w_2}}{\pmatrix{z_2\cr -w_1}} \right ] \right \}\, ,
\end{eqnarray}

where\cite{SJ} 

\begin{equation}
\left [ {\pmatrix{x & y}}{\pmatrix{x'\cr y'}} \right ] = (xy'-yx')\, .
\end{equation}

Define

\begin{eqnarray}
T^{(2)} &=& \left ( g(p',q',r',s',u',v';
z_1,z_2,z_3,w_1,w_2)\psi_0{\bf ,} \quad  g(p,q,r,s,u,v;
z_1,z_2,z_3,w_1,w_2)\psi_0 \right )\nonumber\\ 
&&\nonumber \\
&=& \left (\! {\exp} \left \{\! p'{\pmatrix{r' & s'}}\!\! 
{\pmatrix{z_1\cr w_2}}\! +\! 
q'{\pmatrix{r' & s'}}\!\!{\pmatrix{z_2 \cr -w_1}}\! -\! {\frac{v'}{z_3}} 
\!\left [\! {\pmatrix{z_1 & w_2}}\!\!{\pmatrix{z_2\cr -w_1}}\! \right ]\!
\right \}\psi_0\right. {\bf ,}\nonumber \\
&&\nonumber \\
&&\left. {\exp} \left \{\! p{\pmatrix{r & s}}\!\!{\pmatrix{z_1\cr w_2}}\!
+\!q{\pmatrix{r & s}}\!\!{\pmatrix{z_2 \cr -w_1}}\!
-\!{\frac{v}{z_3}}\left [\! {\pmatrix{z_1 & w_2}}\!\!{\pmatrix{z_2\cr -w_1}}
\!  \right ]\! \right \}\! \psi_0\right )\, ,
\end{eqnarray}
where

\begin{eqnarray}
\frac{\partial}{\partial z_i}\psi_0
=\frac{\partial}{\partial w_i}\psi_0=0\qquad\, , i,j=1,2,3\, .
\end{eqnarray}

To begin computing the above scalar product we have to note that
$z_i, w_i$ and $\frac{\partial }{\partial z_i}$ and
$\frac{\partial }{\partial w_i}$ are adjoints of each
other{\cite{BV}}, for each $i=1,2,3$, under the scalar product.
We also note that adjoints of sums and products are sum of
adjoints and product of adjoints, respectively.

But we will not be required to prove it here since our scalar
product for the basis functions under this assumption reproduces
our previous results.

Therefore,

\begin{eqnarray}
{{\frac{\partial}{\partial p^{\prime}}}T^{(2)}} =
p(rr'+ss')T^{(2)} + \frac{v{\bar v}'}{{\bar
z_3}z_3}{{\frac{\partial}{\partial p^{\prime}}}T^{(2)}}\, ,
\end{eqnarray}

and
\begin{eqnarray}
(1-\frac{v{\bar v}'}{z_3{\bar z_3}}){{\frac{\partial}{{\partial}}
p^{\prime}}}T^{(2)} = p(rr'+ss')T^{(2)}\, .
\end{eqnarray}

Similarly,

\begin{eqnarray}
(1-\frac{v{\bar v}}{z_3{\bar z_3}}){{\frac{\partial}{\partial
q^{\prime}}}T^{(2)}} = q(rr'+ss')T^{(2)}\, .
\end{eqnarray}

The solution of these equations is,

\begin{eqnarray}
T^{(2)}={\exp}\left ( {\frac{(pp'+qq')(rr'+ss')}{1-\frac{v{\bar
v}'}{z_3 {\bar z_3}}}} \right )\times T^{(2)}_0\, ,
\end{eqnarray}

where

\begin{eqnarray}
T^{(2)}_0 &=& \left ( {\exp}\, v\left [ {\pmatrix{z_1 & 
w_2}}{\pmatrix{z_2 \cr -w_1}}
\right ]\psi_0{\bf ,} \quad {\exp}\, v'\left [ {\pmatrix{z_1 &
w_2}} {\pmatrix{z_2 \cr -w_1}}\right ] \psi_0\right )\, ,\nonumber\\
&&\nonumber \\
&=&(\psi_0{\bf ,}\quad Q\psi_0)\, ,
\end{eqnarray}

and

\begin{eqnarray}
Q={\exp}\,( \frac{v}{z_3}\left [
\pmatrix{\frac{\partial}{\partial z_1}   & \frac{\partial}{\partial  w_2}}
\pmatrix{\frac{\partial}{\partial z_2} \cr \frac{\partial}{-\partial w_1}}
\right ] ){\bf\cdot} \quad {\exp}\, \frac{v'}{z_3}\left [ 
\pmatrix{{z_1}   & {w_2}} \pmatrix{{z_2} \cr {-w_1}}\right ] \, .
\end{eqnarray}

To evaluate $T^{(2)}_0$ we follow Schwinger's
procedure\cite{SJ}.  For the convenience of the reader the
derivation of Schwinger fro the general case $T^{(n)}_0$ is given
in the appendix.  We simply quote the result below(see appendix Eq.\ref{T_0}),

\begin{eqnarray}
T^{(2)}_0 &=& \frac{1}{\vert 1 + \kappa{\bar\lambda}\vert }\, ,
\end{eqnarray}

where the vertical bars indicate a determinant.

For us

\begin{eqnarray}
\kappa &=& \pmatrix{0 & \frac{v'}{z_3} \cr -\frac{v'}{z_3} & 0},\nonumber \\
{\bar \lambda} &=& \pmatrix{0 & \frac{\bar v}{\bar z_3} \cr -
\frac{\bar v}{z_3} & 0}\, .
\end{eqnarray}

Therefore

\begin{eqnarray}
\vert 1 + \kappa{\bar \lambda} \vert &=& (1 - \frac{v'{\bar
v}}{z_3{\bar z_3}})^2,\nonumber \\
&& \nonumber \\
T^{(2)}_0 &=& \frac{1}{(1 - \frac{v'{\bar v}}{z_3{\bar z_3}})^2}\, .
\end{eqnarray}

This gives us the scalar product between the generating functions
for the basis  states as 

\begin{eqnarray}
T^{(2)}= \frac{1}{(1 - \frac{v'{\bar v}}{z_3{\bar z_3}})^2}
{\exp}\left ( {\frac{(pp'+qq')(rr'+ss')}{1-\frac{v{\bar
v}'}{z_3 {\bar z_3}}}} \right )\, .
\label{T2}
\end{eqnarray}

As in our previous work, we observe that $\vert z_3\vert^2$
appears in the denominators in Eq.(\ref{T2}).  We
therefore use the same choice for $z_3$ namely set it to,

\begin{eqnarray}
z_3=\exp(i\theta )\, ,
\end{eqnarray}
so that $\vert z_3\vert^2=1$.  

With this choice the scalar product now yeilds

\begin{eqnarray}
T^{(2)}= \frac{1}{(1 - v'{\bar v})^2}
{\exp}\left ( {\frac{(pp'+qq')(rr'+ss')}{1-v'{\bar v}}} \right )\, .
\label{T^{(2)}}
\end{eqnarray}

We still have to specify the measure over the $\theta $
variable.  We keep this as we had done in our previous paper.
That is for us also the scalar product over the angle variable
$\theta$ is an averaging over the angle.  So that,
\begin{eqnarray}
T_\theta =\int^{+\pi}_{-\pi} \frac{d\theta}{2\pi}\exp u'(\exp (-i\theta ))
\exp u(\exp (i\theta
))=\sum^{\infty}_{n=0}\frac{(u'u)^n}{n!^2}\, .
\end{eqnarray}

Therefore taking into account the $T_\theta $ factor we now
write the final expression for the scalar product between the
generating functions for the basis as 

\begin{eqnarray}
T^{(2)} &=& \left ( g(p',q',r',s',u',v';
z_1,z_2,z_3,w_1,w_2)\psi_0{\bf ,} \quad  g(p,q,r,s,u,v;
z_1,z_2,z_3,w_1,w_2)\psi_0 \right )\, ,\nonumber\\ 
&&\nonumber \\
&=&\frac{1}{(1 - v'{\bar v})^2}
{\exp }\left ( {\frac{(pp'+qq')(rr'+ss')}{1-v'{\bar v}}} \right )
\sum^{\infty}_{n=0}\frac{(u'u)^n}{n!^2}\, . \nonumber \\
\label{NSP}
\end{eqnarray}

Similarly the scalar product between two generalized generating
functions is 
\begin{eqnarray}
{(\cal{G'}, \cal{G})} &=& (1 - v'v)^{-2} 
{\exp}\left [ (1 - v'v)^{-1}({r_p}'r_p +
{r_q}'r_q + {s_p}'s_p + {s_q}'s_q) \right ]\nonumber\\
&&\nonumber \\
&&\times \sum^{\infty}_{n=0}\frac{1}{n!^2}
\left ( u'- v\frac{({r_p}'{s_q}'
+ {r_q}'{s_p}')}{(1 - v'v)}\right )^n
\, {\bf \cdot}\,\left (u - v'\frac{({r_p}{s_q} 
+ {r_q}{s_p})}{(1 - v'v)}\right )^n\, .
\end{eqnarray}

\setcounter{equation}{0}
\section{Computation of the Clebsch-Gordan  coefficients using
the 'differential measure'}

There are sevaral methods of computing the Clebsch-Gordan
coefficients of $SU(3)$.  One of these is the method of vector
invariants first used by van der Waerden to compute the
Clebsch-Gordan coefficients for $SU(2)$.  In this way of
computing these coefficients one has to first write down the
most general vector invariant of $SU(3)$ in terms of the basis
states of three pairs of ${\underline 3}$ and ${\underline 3^*}$
of $SU(3)$.  This invariant may be expanded in terms of product
of basis staes of three irreducible representations of $SU(3)$
and the expansion coefficients are the required Clebsch-Gordan
coefficients of $SU(3)$.  We borrow from our previous
paper\cite{JSPHSS} the expression for the generating function
for the most general genaral vector invariants of $SU(3)$
Eq.(\ref{GFI}).  We recall here that the basis states of $SU(3)$
are also available to us in the form of a generating function
(\ref{g()}).  We extract the Clebsch-Gordan coefficients by
making use of 'auxiliary scalar product'.  In view of this the
computation of the Clebsch-Gordan coefficients of $SU(3)$ reduces
to the problem of computing the scalar product,
\begin{eqnarray}
\int_{\pm}=\left (g^1g^2g^3,\,\, {\cal{I}_{\pm}}\right )\, .
\end{eqnarray}
where $g^a$ is the generating function for the basis states in
the variables $z^a_i,\, w^a_i,\,\, a,i=1,2,3$ and ${\cal{I_+}}$
is the generating function for the invariants involving the
trilinear invariant in the $z^a_i$ whereas ${\cal{I}_-}$ is the
generating function for the invariants involving the trilinear
invariant in the $w^a_i$.  Throghout the following the upper
index denotes one of the three IR's and the lower index denotes
the components of the vectors in the give IR.  It is enough for
us to compute the scalar product involving ${\cal{I}_+}$ the
other one can then be deduced from this by a simple
argument\cite{JSPHSS}.

\subsection{Evaluation of $\int_+$ - The $\theta$ Variable Part}

We now evaluate $\int_+$.  We have,
\begin{eqnarray}
\int_+=\left (g^1g^2g^3,\,\, {\cal{I}_+} \right )\, .
\end{eqnarray}

Here,
\begin{eqnarray}
g^a=\exp \left
(r^a_pz^a_1+s^a_qw^a_1+s^a_pw^a_2+u^ae^{i\theta^a}-
v^ae^{-i\theta^a}(z^a_1w^a_1+z^a_2w^a_2)\right )\nonumber \\
{\mbox{a=1,2,3}}\, .
\label{g^a}
\end{eqnarray}

In Eq.(\ref{g^a}) we have to use the meaning of the variables
implied by the Eq.(\ref{s_q=-sq}).  Also, in ${\cal{I}_+}$ we
have, 
\begin{eqnarray}
{\vec{z}^a}\cdot {\vec{w}^a}=z^a_1w^b_1+z^a_2w^b_2-\exp 
(i\theta^a-i\theta^b)(z^b_1w^b_1+z^b_1w^b_2)\, ,
\end{eqnarray}
and
\begin{eqnarray}
{\vec{z^1}}\cdot {\vec{z^2}}\times
{\vec{z^3}}=e^{i\theta^1}(z^2_1z^3_2-z^3_1z^2_2)+({\mbox{cyclic}})
\, .
\end{eqnarray}

We have no new way of evaluating the $\theta$ integrations
required to be done in the above scalar product.  Hence for this
part we borrow the results of\cite{JSPHSS}.  

\subsection{Evaluation of $\int_+$ - The $z^a_i$ and $w^a_i$
Part, where $a=1,2,3,\,{\mbox{ and }} i=1,2$}

Next we use this scalar product to extract the Clebsch-Gordan
coefficients of $SU(3)$ from the remaining part of the most
general expression for the invariants of $SU(3)$ formed out of
the pair of vectors $z^i_j$ and $w^i_j$ where $i,j=1,2,3$.  The
upper index denotes the IR and the lower one denotes the
component in each IR.

The remaining part of the scalar product after doing the $\theta$
integrations is denoted by $T$ and is given below.

\begin{eqnarray}
T=\left ({\exp}\left\{\sum_{i,j}-V^{ij} (z^i_1w^i_1 + z^i_2w^i_2)+
p^i(r^iz^i_1 + s^iw^i_2) + q^i(r^iz^i_2 - s^iw^i_1)\right. \right
\}\psi_0 ,\nonumber \\
\left. {\exp}\left \{\sum_{m,n}-J_{mn}(z^m_1w^n_1 + z^m_2w^n_2) +
j_+z^m_1A_{mn}z^n_2\right \}\psi_0 \right )\, .
\label{ellipses}
\end{eqnarray}
where
\begin{eqnarray}
A=\pmatrix{0 & 1 & -1 \cr -1 & 0 & 1 \cr 1 & -1 & 0}\, .
\end{eqnarray}

Next differentiating $T$ with respect to $p_i$ we get

\begin{eqnarray}
\frac{\partial T}{\partial p^i} &=& \left ( {\exp}(\cdots )\psi_0,
\qquad (r^i\frac{\partial }{\partial z^i_1} + s^i\frac{\partial
}{\partial w^i_2}){\exp}(\cdots )\psi_0\right )\, ,\nonumber \\
&&\nonumber\\
&=&\left ({\exp}(\cdots )\psi_0, \left \{\sum_{m,n} - r^i J_{mn} w^n_1
\delta_{mi} - s^i J_{mn} z^m_2 \delta_{ni} + j_+r^i A_{mn} z^n_2 
\delta_{mi}\right \} {\exp}(\cdots )\psi_0\right )\, ,\nonumber
\\ &&\nonumber\\
&=&\left (\left \{\sum_n -r^i J_{in} {\frac{\partial}{\partial w^n_1}}
- \sum_m s^i J_{mi} \frac{\partial}{\partial z^m_2}
+ j_+\sum_n r^i A_{in} \frac{\partial}{\partial
z^n_2}{\exp}(\cdots ) \right \}\psi_0,
{\exp}(\cdots)\psi_0\right ) \, ,\nonumber \\
\end{eqnarray}
where the ellipses for the arguments in the $\exp$ function
stand for the same arguments of the $\exp $ functions in the
Eq.(\ref{ellipses}) above.

We will first evaluate the effect of the term involving $A$. So let,

\begin{eqnarray}
T'=\sum_n r^iA_{in}\left (\frac{\partial}{\partial z^n_2}
{\exp}(\cdots )\psi_0{\bf ,}
\qquad {\exp}(\cdots)\psi_0\right )\, .
\end{eqnarray}

This gives us,

\begin{eqnarray}
\sum_m \left(1-VJ^T)_{mn} \right ) T' = \left (\sum_a r^i A_{ia}
r^a_q\right ) T\, .
\end{eqnarray}

Therefore

\begin{eqnarray}
\frac{\partial T}{\partial p^i}
&=&\left (\left \{\sum_n -r^i J_{in} {\frac{\partial}{\partial
w^n_1}} - \sum_m s^i J_{mi} \frac{\partial}{\partial
z^m_2}{\exp}(\cdots )\right \}\psi_0, {\exp}(\cdots)\psi_0\right
) + j_+T'\, ,\nonumber\\
&&\nonumber\\
&=&\left (\left \{\sum_n -r^i J_{in} {\frac{\partial}{\partial
w^n_1}} - \sum_m s^i J_{mi} \frac{\partial}{\partial
z^m_2}{\exp}(\cdots )\right \}\psi_0, {\exp}(\cdots)\psi_0\right )\nonumber\\
&&+ j_+r^i\left (\sum_{m,n} A_{im} \frac{1}{(1-VJ^T)_{mn}} r^n_q
\right )T\, ,\nonumber \\
&&\nonumber\\
&=&  j_+r^i\left (\sum_{m,n} A_{im} \frac{1}{(1-VJ^T)_{mn}} r^n_q \right )T
-\left ( \sum_{a} (r^i J_{ia} s^a_q + s^i J_{ai} r^a_q)\right ) T\nonumber \\
&&+ \sum_{a} (VJ^T)_{ai}\frac{\partial T}{\partial p^a}\, .
\end{eqnarray}

Hence
\begin{eqnarray}
\frac{\partial T}{\partial p^i}
\!- \!\sum_{a,b} V^{ab}J^T_{bi}\frac{\partial T}{\partial p^a}
 &=&  \sum_{a,b} \left (j_+r^iA_{ia} \frac{1}{(1-VJ^T)_{ab}} r^b_q 
\!-\! r^i J_{ia} s^a_q \!-\! s^i J_{ai} r^a_q\right ) T\, .
\end{eqnarray}

Expressed in matrix notation, the solution of this equation is 

\begin{eqnarray}
T={\exp}\left [
-S^T_qJ^T(1-VJ^T)^{-1}R_p-S^T_pJ^T(1-VJ^T)^{-1}R_q\right. \nonumber \\
\left. +j_+R^T_qJ^T(1-VJ^T)^{T^{-1}}A^T(1-VJ^T)^{-1}R_p\right ] T^{(6)}_0\, ,
\end{eqnarray}

where 
\begin{eqnarray}
R_p=\pmatrix{r^1_p \cr r^2_p \cr r^3_p}\, ,
\end{eqnarray}
and similar notation follows for $R_q, S_p, S_q$.

In the above

\begin{eqnarray}
T^{(6)}_0&=&\left (\left \{ {\exp}\sum_{i,j}-V'^{ij} (z^i_1w^i_1
+ z^i_2w^i_2)\right \} \psi_0,\quad 
\left \{ {\exp}\sum_{m,n}-J'_{mn}(z^m_1w^n_1 + z^m_2w^n_2) +
j_+z^m_1A_{mn}z^n_2\right \}\, \psi_0\right )\, ,\nonumber \\
&&\nonumber\\
&=&\left (\left \{ {\exp}\sum^6_{\mu ,\nu = 1}V'^{\mu\nu}[A_\mu
A_\nu] \right \}\psi_0,\quad 
\left \{ {\exp}\sum^6_{\mu,\nu = 1}J'_{\mu\nu}[A_\mu A_\nu
]\right \} \psi_0\right )\, ,
\label{T^6_0}
\end{eqnarray}
where
\begin{eqnarray}
A_1={\pmatrix{z^1_1 & z^1_2}},\quad A_2={\pmatrix{w^1_2 \cr -w^1_1}},
\quad A_3={\pmatrix{z^2_1 & z^2_2}},\nonumber \\
A_4={\pmatrix{w^2_2 \cr -w^2_1}},\quad A_5={\pmatrix{z^3_1 & z^3_2}}, 
\quad A_6={\pmatrix{w^3_2 \cr -w^3_1}}\, ,
\end{eqnarray}

\begin{eqnarray}
[xy] = x_1y_2 - x_2y_1\, , \,\,\, \mbox{with}\,\, x=\pmatrix{x_1
\cr x_2}\,\, \mbox{and}\,\, y=\pmatrix{y_1 & y_2}\, ,
\label{[]}
\end{eqnarray}

\begin{eqnarray}
V'&=&\pmatrix{0 & v^{12} & 0 & 0 & 0 & 0 \cr - v^{12} & 0 & 0 & 0 & 0 & 0 
        \cr 0 & 0 & 0 & v^{34} & 0 & 0 \cr 0 & 0 & -v^{34} & 0 & 0 & 0 
\cr 0 & 0 & 0 & 0 & 0 & v^{56} \cr 0 & 0 & 0 & 0 & -v^{56} &
0}\, ,\nonumber \\
&&\nonumber \\
&&\nonumber \\
&=&\pmatrix{0 & v^{1} & 0 & 0 & 0 & 0 \cr - v^{1} & 0 & 0 & 0 & 0 & 0 
        \cr 0 & 0 & 0 & v^{2} & 0 & 0 \cr 0 & 0 & -v^{2} & 0 & 0 & 0 
\cr 0 & 0 & 0 & 0 & 0 & v^{3} \cr 0 & 0 & 0 & 0 & -v^{3} & 0}\, ,
\nonumber\\
\nonumber\\
J'&=&\pmatrix{0 & J_{12} & j_+ & J_{14} & j_+ & J_{16}   \cr 
           -J_{12} & 0 & J_{32} & 0 & J_{52} & 0      \cr 
           -j_+ & -J_{32} & 0 & J_{34} & j_+ & J_{36} \cr 
           -J_{14} & 0 & -J_{34} & 0 & J_{54} & 0     \cr 
           -j_+ & -J_{52} & -j_+ & -J_{54} & 0 & J_{56} \cr 
           -J_{16} & 0 & -J_{36} & 0 & -J_{56} & 0}\nonumber\\
&&\nonumber \\
&&\nonumber \\
&=&\pmatrix{0 & j_{31}+j_{21} & j_+ & j_{12} & j_+ & j_{13}   \cr 
           -(j_{31}+j_{21}) & 0 & j_{21} & 0 & j_{31} & 0      \cr 
           -j_+ & -j_{21} & 0 & j_{12}+j_{32} & j_+ & j_{23} \cr 
           -j_{12} & 0 & -(j_{12}+j_{32}) & 0 & j_{32} & 0     \cr 
           -j_+ & -j_{31} & -j_+ & -j_{32} & 0 & j_{23}+j_{13} \cr 
           -j_{13} & 0 & -j_{23} & 0 & -(j_{23}+j_{13}) & 0}\, .\nonumber \\
\end{eqnarray}

We borrow the expression for $T^{(6)}_0$ from Schwinger (see
appendix Eq.(\ref{T_0})).  For us $\kappa = J', \lambda = V'$
and, as before, we will assume that both $J'$ and $V'$ are real
matrices as this makes no difference to our results.

With this identification we get,

\begin{eqnarray}
T^{(6)}_0&=&\frac{1}{\vert 1+J'V' \vert }\, ,\nonumber\\
&=&\frac{1}{\vert 1+(V')^T(J')^T \vert }\, ,\nonumber \\
&=&\frac{1}{\vert 1-(V')(J')^T \vert }\, .
\end{eqnarray}

By expanding this determinant one can show that

\begin{eqnarray}
T^{(6)}_0&=&\frac{1}{\vert 1-V'(J')^T \vert }\, ,\nonumber \\
&& \nonumber \\
&=&\frac{1}{\left (\vert 1 - VJ^T\vert \right )^2}\, .
\end{eqnarray}
where $V, J$ are as defined in Eq.(\ref{V&J}).

Putting together our results obtained so far we see that,

\begin{eqnarray}
\int_+ &=&{\exp}\left [
-S^T_qJ^T(1-VJ^T)^{-1}R_p-S^T_pJ^T(1-VJ^T)^{-1}R_q\right. \nonumber \\
&&+\left. j_+R^T_qJ^T(1-VJ^T)^{T^{-1}}A^T(1-VJ^T)^{-1}R_p\right
] T^{(6)}_0\, ,\nonumber \\
&=&\frac{1}{(\vert 1-VJ^T\vert)^2}{\exp}\left [ -S^T_qJ^T(1-VJ^T)^{-1} 
R_p-S^T_pJ^T(1-VJ^T)^{-1}R_q\right.\nonumber \\
&&+\left. j_+R^T_qJ^T(1-VJ^T)^{T^{-1}}A^T(1-VJ^T)^{-1}R_p\right ]\, .
\label{GFCG}
\end{eqnarray}

We note that this is the expression that we got for $\int_+$ in
our paper\cite{JSPHSS}.  The expression for $S\int_-$, which
corresponds to the case $L<0$ can be got from the one for
$\int_+$ easily by a simple argument (see \cite{JSPHSS}).

\subsection{Overall normalization factor of the Clebsch-Gordan coefficients}

In our erlier work\cite{JSPHSS} we obtained a formula for the
Clebsch-Gordan coefficients of $SU(3)$ upto an overall
normalization factor, which we denoted by $n(N(1,2),
N(1,3),\cdots )$Eq.(\ref{overall}).  This factor depends on the
three IRs, for which the Clebsch-Gordan coefficients are being
calculated, only and not on the individual basis states. It
corresponds to the norm of the most general invariant\cite{BV}
constructed out of three arbitrary IRs of $SU(3)$.  It is a
function of the multiplicity labels $N(1,2,\cdots , L$. We now
wish to compute the same with help of the tools developed so
far.  To do this we proceed as before, namely we compute the
norm of the generating function, for the invariants, itself.
This will give us a generating function for the normalization
factors $n(N(1, 2)\cdots )$.  To do this we first have to
evaluate the $\theta $ part of the scalar product.  We recall
that on erlier occasions we have not handled this using our
'differential' measure.  We therefore propose to use, as in our
previous work\cite{JSPHSS}, an averaging over the $\theta $
variable as the scalar product for that part of the of the
scalar product.  Now by mimicking the argument in that paper for
doing the $\theta$ integrations word to word we come to the same
conclusion, as in that paper, that is that we need not do these
$\theta $ integrations actually but can actually drop them.
Therefore with this understanding we write the resulting scalar
product over the remaining complex variables $z^i_j, w^i_j$,
where $i=1,2,3$ and $j=1,2$, as below.

\begin{eqnarray}
T^{'}_0&=&\left ({\exp}\left \{\sum_{m',n'} -
J^{'}_{m'n'}(z^{m'}_1w^{n'}_1 + z^{m'}_2w^{n'}_2) 
+ j^{'}_+ z^{m'}_1 A_{m'n'} z^{n'}_2\right \}\psi_0 \right. ,\nonumber \\
\nonumber \\
&&\left. {\exp}\left \{\sum_{m,n}-J_{mn}(z^m_1w^n_1 + z^m_2w^n_2) +
j_+z^m_1A_{mn}z^n_2\right \}\psi_0 \right )\nonumber \\
\nonumber \\
&=&\left (\left \{ {\exp}\sum^6_{\mu ,\nu = 1}J"^{\mu\nu}[A_\mu A_\nu]\right \}\psi_0,\quad 
\left \{ {\exp}\sum^6_{\mu,\nu = 1}J'_{\mu\nu}[A_\mu A_\nu
]\right \}\psi_0\right )\, ,\nonumber \\
\nonumber \\
&=&\frac{1}{\vert 1 + J" J' \vert }\, .
\label{JJ}
\end{eqnarray}
where, following the notation in Eq.(\ref{T^6_0}), $J^{'}$ is a $6
\time 6$ matrix whose elements are functions of $j_{12}, \cdots
,j_+, L$ and $J^{''}$ is a similar $6 \times 6$ matrix whose
elements are functions of $j^{'}_{12}, \cdots ,j^{'}_+, L^{'}$.  The last
result follows from Eq.(\ref{T_0}).

Formally we can write

\begin{eqnarray}
T^{'}_0 &=& \sum^\infty_{N^{'}(1, 2),\, N^{'}(1, 3),\, N^{'}(2,
1),\, N^{'}(3, 1),\, N^{'}(2, 3),\, N^{'}(3, 2),\, j^{'}_+,\,\,
N(1, 2),\, N(1, 3),\, N(2, 1),\, N(3, 1),\, N(2, 3),\, N(3,
2),\, j_+\, =\, 0} n(N(1, 2), N^{'}(1,2) \cdots )\nonumber \\
\nonumber \\
&&\times {j^{'}}^{N^{'}(1,2)}_{12} {j^{'}}^{N^{'}(1,3)}_{13}
 {j^{'}}^{N^{'}(2,1)}_{21}
{j^{'}}^{N^{'}(3,1)}_{31} {j^{'}}^{N^{'}(2,3)}_{23}
{j^{'}}^{N^{'}(3,2)}_{32} j^L_+j^{N(1,2)}_{12} j^{N(1,3)}_{13}
j^{N(2,1)}_{21} j^{N(3,1)}_{31} j^{N(2,3)}_{23} j^{N(3,2)}_{32}
j^L_+ \, .\nonumber\\
\label{n()}
\end{eqnarray}

Therefore to get the normalization factor $n(N(1,2)\cdots )$ we
have to expand the right hand side of Eq.(\ref{JJ}) in a taylor
series expansion in the monomials $(j^{'}_{12} j_{12})^{N(1,2)}
 (j^{'}_{13}j_{13})^{N(1,3)} \cdots (j^{'}_+j_+)^L$ and extract
its coefficient.

\section{Discussion}

Making use of Eq.(\ref{GFCG}) one can derive an explicit closed
form algebraic expression for the Clebsch-Gordan coefficients of
$SU(3)$(Eq.\ref{AECG}).  This is done by computing the scalar
product between the invariants and basis states in two ways, (i)
using the formal expressions and, (ii) using the explicit
realizations.  By equating these two expressions the formula for
the Clebsch-Gordan coefficient is obtained.  The derivation is
given in detail in \cite{JSPHSS} to which the interested reader
is referred.

In this paper, which is devoted mostly to rederive our previous
results, we have computed the overall normalization $n(\cdots )$
(Eq.\ref{n()}) using our 'auxiliary differential measure'.  We
recall that this was left uncomputed in our paper using Gaussian
integration techniques.  In this task as in other tasks
Schwinger's formula has been of great utility to us.

Now let us note a few facts regarding our 'differential measure'
for $SU(3)$.  

Schwinger\cite{SJ} while trying to construct basis states of the
total angular momentum operator used a scalar product.  It is a
remarkable fact that the scalar product in both cases turn out
to be almost identical to each other though, in the case of
Schwinger it provides the actual normalizations for the basis
states whereas in our case it provides only auxiliary
normalizations.  It is only almost exact because in our,
$SU(3)$, case we have the scalar product between basis functions
which are dependent on four complex variables and a planar rotor
variable.  In the case of Schwinger there are only four complex
variables and for him the scalar product does not involve any
other variable.  Therefore we in our, $SU(3)$, case have an
additional piece, coming from the planar rotor varible part,
added to our scalar product.  Compare Eq.(C7) of (\cite{SJ}) with
Eq.(\ref{gg'}).

The just mentioned relation between the scalar products of the
basis functions of the total angular momentum operator and those
of the group $SU(3)$ raises the possibilty of the
corresopsonding basis functions being related to each other.  In
fact we can relate every basis state of $SU(3)$, Eq.(\ref{g()}),
with every basis state of $SU(2) \times SU(2)$ (Eq.(3.35) of
\cite{SJ})(or equivalently with that of $SO(1,3)$ (only finite 
dimensional IRs in this case) and vice versa in a one-to-one
fashion.  This kind of duality is known already as a general
principle.  But our model space plus generating function method
of studying group reprsentations has made it possible to realize
this duality in a simple and concrete way.  A particular use of
this duality may be the possibilty of interpreting the IR and
multiplicity labels of $SU(3)$ in terms of angular momentum
quantum numbers.  Investigations of this nature will form a part
of a future publication.\footnote{manuscript under preparation}

Ruegg\cite{RH} in his work on the Clebsch-Gordan coefficients of
$SU_q(2)$ used a 'differential measure' to derive the correct
normalizations for the basis states and later the $SU_q(2)$
Clebsch-Gordan coefficients.  For him it was a necessity since
Quantum groups are not groups in the same sense as Lie groups
and therefore a group invariant measure and a scalar product in
terms of this measure do not exist.  So he first defined a
'differential measure' for $SU(2)$ and then extended it to
$SU_q(2)$ by making use of the $q$-derivatives, $q$- factorials,
and $q$-binomial theorem etc.  It is also clear now that the
extension of our results to the group $SU_q(3)$ can proceed on
lines similar to $SU_q(2)$.  In principle one should be able to
mimic the procedure of Ruegg designed for $SU_q(2)$ for the
group $SU_q(3)$ also using the realization of $SU_q(3)$ algebra
in terms of four pairs of boson operators and a planor rotor
operator which we obtained in our erlier work(see appendix B of
\cite{JSPHSS})

Another interesting fact to note is that when one tries extend
our results for $SU(3)$ to groups $SU(n), n > 3$ one has to
grapple with two problems.  One is the construction of a
generating function for the basis states and the other is the
construction of a measre to be used in the scalar product
between the basis states.  The nature of the 'differential
measure' makes it easily adaptable to generalization though as
one can see clealry the algebra that one has to go through to
compute the scalar products is a little more tedious in the case
of the 'differential measure' than in the case of Gaussian
integrations.\\

Finally we note that since the present calculus is in complete
analogy with that of Schwinger for $SU(2)$\cite{SJ} we can try
to apply his techniques of obtaining various
objects\cite{JSPWEYL,JSPWIGD} on $SU(3)$.

\noindent {\large{\bf{Acknowledgements}}}\\

I wish to thank Koushik Ray for the discussions we have had
during this work.  Part of the work was done when the author was
visiting Centre for Theoretical Studies(CTS), IISc, Bangalore
during the year 1994. He wishes to thank CTS for the
hospitality.\\

\renewcommand{\theequation}{A.\arabic{equation}}
\setcounter{equation}{0}

\newpage

{\large\bf Appendix : Schwinger's derivation of the scalar 
product $T^{({n})}_0$}

Schwinger evaluated the scalar product $T^{(n)}_0$(see Appendix
C of \cite{SJ}).  For convenience we reproduce his derivation
here almost verbatim.

For this purpose let,

\begin{eqnarray}
T^{(n)}_0 = \left (\psi_0, Q\psi_0\right )\, ,
\end{eqnarray}

where 
\begin{eqnarray}
Q={\exp}\left (\frac{1}{2}\sum^n_{\mu ,\nu = 1}(\lambda^*)_{\mu\nu}
[A_\mu A_\nu]\right ){\bf{\cdot}}\quad
{\exp}\left (\frac{1}{2}\sum^n_{\mu,\nu = 1}\kappa_{\mu\nu}
[A^\dagger_\mu A^\dagger_\nu ]\right )\, .
\end{eqnarray}
and the square brackets $[]$ have the same meaning as in
Eq.(\ref{[]}).  The $A_\mu$ are $n$ sets of two-component
operators obeying

\begin{eqnarray}
[A_{\zeta\mu},\, A^\dagger_{\zeta^{'}\nu} ]
=\delta_{\mu\nu}\delta_{\zeta\zeta^{'}}\,.
\end{eqnarray}
while $\lambda_{\mu\nu}$ and $\kappa_{\mu\nu}$ form
antisymmetrical matrices.\footnote{This fact is not used in the proof}

We note the following properties of $Q$

\begin{eqnarray}
\left (\frac{\partial Q}{\partial \lambda^*_{\mu \nu}}\right ) 
&=& [A_\mu A_\nu ]Q, \\
\left [ [xA_\mu ], \quad Q\right ] &=& -Q\sum_\nu \kappa_{\mu
\nu} \left (xA\right ),\\
\left [ Q, \quad (xA^\dagger_\mu ) \right ] &=& \sum_\nu 
\lambda^*_{\mu \nu}[xA_\nu ]Q\, ,
\end{eqnarray}
in which $X$ is an arbitrary constant spinor and the left hand
sides of the last two equations are commutators. 
The above equations can be combined to give

\begin{eqnarray}
\sum_\nu (1+\kappa\lambda^*)_{\mu \nu}[xA_\nu]Q = Q[xA_\nu]Q -
\sum_\beta \kappa_{\mu \beta}
(xA^\dagger_\beta)Q,\\
{\mbox{or}}
\qquad [xA_\nu]Q = \sum_{\nu\beta}\left (
\frac{1}{1+\kappa\lambda^*} \right )_{\nu\beta} 
Q[xA_\beta] - \sum_{\nu\beta}\left ( \frac{1}{1+\kappa\lambda^*}
\right )_{\nu \beta}
(xA^\dagger_\beta)Q\, .
\end{eqnarray}

Therefore

\begin{eqnarray}
[A_\mu A_\nu ]Q=\sum_\beta\left (
\frac{1}{1+\kappa\lambda^*}\right )_{\nu\beta} [A_\mu QA_\beta]
-\sum_\beta\left ( \frac{1}{1+\kappa\lambda^*}\right
)_{\nu\beta} (A^\dagger_\beta A_\mu )Q\nonumber \\
-2\left (\frac{1}{1+\kappa\lambda^*}\kappa\right )_{\nu\mu}Q\, ,
\end{eqnarray}

from which we obtain, since $\left (A_\mu \psi_0 = 0\right )$

\begin{eqnarray}
\left ( \frac{\partial}{\partial \lambda^*_{\mu\nu}}\right )T^{(n)}_0 
= -2\left ( \frac{1}{1+\kappa\lambda^*}J\right )_{\nu\mu}T^{(n)}_0\, .
\end{eqnarray}

Thus with respect to changes in the matrix $V'$, we have

\begin{eqnarray}
\delta {\mbox{log}} T^{(n)}_0 = \frac{1}{2}\sum_{\mu\nu}\delta 
\lambda^*_{\mu\nu} 
(\frac{\partial }{\partial \lambda^*_{\mu\nu}}){\mbox{log}}
T^{(n)}_0 =-{\mbox{tr}}
\left ( \frac{1}{1+\kappa\lambda^*}\kappa\delta \lambda^*\right
) \, .\nonumber \\
\end{eqnarray}

On comparing this with the theorem on differentiation of a 
determinant\cite{MJRNH},

\begin{eqnarray}
\delta {\mbox{log}}\vert M \vert = {\mbox{tr}}(M^{-1}\delta M )\, ,
\end{eqnarray}

we obtain the desired general result,

\begin{eqnarray}
T^{(n)}_0=\frac{1}{\vert 1+\kappa\lambda^* \vert }\, ,\nonumber\\
\label{T_0}
\end{eqnarray}
where the vertical bars indicate a determinant.

\end{document}